\documentclass[fleqn]{article}
\usepackage{Gc,epsf}
\usepackage[english]{babel}

\def\I{{\mathcal{I}}}
\def\J{{\mathcal{J}}}

\def\L{{\mathcal{L}}}

\def\p{\partial}

\heads{Alexander A. Chernitskii}
{Gravitation and electromagnetism in theory of a unified four-vector field}

\begin{document}
\twocolumn[
\Arthead{12}{2006}{2-3 (46-47)}{130}{132}

\Title{Gravitation and electromagnetism \yy
       in theory of a unified four-vector field}

\Author{Alexander A. Chernitskii\foom 1}{A. Friedmann Laboratory for Theoretical Physics,
St.-Petersburg, Russia\\ State University of Engineering and Economics,
Marata str. 27, St.-Petersburg, Russia, 191002}

\Abstract
    {A four-vector field in flat space-time, satisfying a gauge-invariant set of
    second-order differential equations, is considered as a unified field. The model variational principle
    corresponds to the general covariance idea and gives rise to nonlinear Born-Infeld electrodynamics.
    Thus the four-vector field is considered as an electromagnetic potential.
    It is suggested that space-localized (particle) solutions of the nonlinear field model correspond to
    material particles. Electromagnetic and gravitational interactions between field particles appear
    naturally when a many-particle solution is investigated with the help of a perturbation method.
    The electromagnetic interaction appears
    in the first order in the small field of distant particles. In the second order, there is an effective Riemannian space
    induced by the field of distant particles. This Riemannian space can be connected with gravitation.}
%\selectlanguage{russian}
%\RAbstract
%    {Гравитация и электромагнетизм
%     в теории единого четырёхвекторного поля}
%    {Александр А. Черницкий}
%    {В качестве единого поля рассматривается четырёхвекторное поле в плоском пространстве-времени,
%удовлетворяющее калибровочно инвариантной системе нелинейных дифференциальных уравнений второго порядка.
%Вариационный принцип модели соответствует идее общей ковариантности и приводит
%к нелинейной электродинамике Борна-Инфельда. Таким образом, четырёхвекторное поле имеет смысл
%электромагнитного потенциала.
%Полагается, что материальным частицам соответствуют пространственно-локализованные (корпускулярные) решения
%рассматриваемой нелинейной полевой модели. Электромагнитное и гравитационное взаимодействия между полевыми частицами
%появляются естественно при исследовании многочастичного решения модели методом возмущений.
%В первом порядке по слабому полю удалённых частиц
%появляется электромагнитное взаимодействие.
%Во втором порядке имеется эффективное Риманово пространство, индуцированное полем удалённых частиц.
%Это Риманово пространство может быть связано с гравитацией.}
] %%%%%%%%%%%%%%  End of temporary one-column mode %%%%%%%%%%%%%%
\email 1 {AAChernitskii@eltech.ru}

%\selectlanguage{english}

\section{Introduction}

The problem of unification of all interactions of material particles is one of the
most important problems in modern theoretical physics. Specifically, unification for
the two known long-range interactions, viz., electromagnetism and gravitation, can be considered as
a priority problem in theoretical investigation of nature.

The approach to this problem presented here is connected with a consistent application of the idea of nonlinear
local unified field theory.

Spatially, localized solutions in this theory correspond to solitary material particles.
Such solutions are said to be solitary or solitonic. Also the term ``soliton'' is used in this case.
These solutions can be designated particle solutions.
But the many-particle world configuration corresponds to a complicated many-particle (many-soliton) solution.
The many-particle solution contains the appropriate particle solutions in the following sense.

Each particle solution has at least ten free parameters for space-time rotation and shift.
Because of the nonlinearity, a sum of particle solutions is not a solution for the model.
But we can consider the free parameters of particle solutions to be weakly time-dependent.
This method is well known in the theory of nonlinear dynamics.
A sum of particle solutions with time-dependent free parameters can be considered
as an initial approximation to a many-particle solution. The time dependence of the free parameters
of the particle solutions corresponds to the interaction between the particles.

This method, applied to a suitable model, must give electromagnetic and gravitational interactions for
the case when the interacting particles are sufficiently distant from each other.

For the first time, this approach to the problem of unification of the gravitational and electromagnetic interactions
appeared in the context of some nonlinear electrodynamics model \cite{Chernitskii1992}.
Later on, the approach was developed for another nonlinear electrodynamic model (Born-Infeld) \cite{Chernitskii1999}.

A distinguishing characteristic of this approach is that the gravitational interaction must appear through
an effective Riemannian space for propagation of particle under consideration.
This effective Riemannian space is induced by the electromagnetic field of distant particles.

\section{Model field and equations}

As mentioned above, the model field is considered to be electromagnetic, satisfying nonlinear equations.
But the model to be considered is an unusual electrodynamics not only because the appropriate equations are
nonlinear. This model does not contain the postulated trajectory equation for charged particle in external
electromagnetic field, i.e., the electromagnetic interaction. This interaction appears naturally as a
manifestation of the nonlinearity of the model.

Thus the model field can be called simply an
antisymmetric space-time tensor field instead of the electromagnetic one. More precisely, the model field is
the space-time vector field (four-vector potential) because it is this field that is varied in the model
variational principle.

The variational principle considered here is similar to the one proposed by A.S. Eddington  in the context of
general relativity ideas \cite{Eddington1924} and investigated by A.~Einstein in the context of his unified field
theory \cite{Einstein1923dE}.
Afterwards, M.~Born and L.~Infeld used it in the context of nonlinear electrodynamics \cite{Born1934,BornInfeld1934a}.
The variational principle has the form
\begin{equation}
\label{70801375}
\delta\int\!\!\sqrt{|\det(G_{\mu\nu})|}\;(\mathrm{d}x)^{4} = 0
\;\;,
\end{equation}
where
\begin{equation}
\label{70959108}
G_{\mu\nu} = g_{\mu\nu} + \chi^2\,F_{\mu\nu}
\;\;,\quad
F_{\mu\nu} = \frac{\partial A_{\nu}}{\partial x^{\mu}} - \frac{\partial A_{\mu}}{\partial x^{\nu}}
\;\;,
\end{equation}
$g_{\mu\nu}$ being the components of the space-time metric tensor, $A_{\mu}$ the components of the
electromagnetic four-potential, and the Greek indexes take the value $0,1,2,3$.

The model based on the principle (\ref{70801375}) is related to Einstein's idea of a nonsymmetrical metric
\cite{ChernikovShavokhina1986} and to minimal surfaces \cite{Shavokhina1989}. In the case of a
consideration of minimal surfaces, the components $A_{\mu}$ contained in (\ref{70959108}) are considered to be
four-vector components for special coordinate transformations only. But for general coordinate transformations
they must be considered as additional four coordinates for determination of a minimal hypersurface. Thus, in this case,
eight-dimensional space is considered.

The variational principle (\ref{70801375}) gives the following set of equations \cite{Chernitskii1998a}:
\begin{equation}
\label{35357121}
\frac{\p}{\p x^\mu}\,\sqrt{|g|}\; f^{\mu\nu} {}={} 0
\;\;,
\end{equation}
where
\begin{equation}
\label{65713937}
f^{\mu\nu} {}\equiv {}
%\frac{1}{\alpha^2}\,
\frac{\chi^{-2}\,\p\L}{\p(\p_\mu A_\nu)} {}={}
\frac{1}{\L}\,\left(F^{\mu\nu}  {}-{}
\frac{\chi^2}{2}\,\J\,\varepsilon^{\mu\nu\sigma\rho}\,F_{\sigma\rho}\right)
,\;
\end{equation}
$\L {}\equiv{}   \sqrt{|\,1 {}-{}  \chi^2\,\I  {}-{}  \chi^4\,\J^2\,|}$,
\mbox{$\I {}\equiv{}  F_{\mu\nu}\,F^{\nu\mu}/2$},\\
\mbox{$\J {}\equiv{}  \varepsilon_{\mu\nu\sigma\rho}\, F^{\mu\nu} F^{\sigma\rho}/8$},
%$\varepsilon_{\mu\nu\sigma\rho}   \equiv  \pm\sqrt{|g|}$,
$\varepsilon_{0123} \equiv \sqrt{|g|}$,
%$\varepsilon^{\mu\nu\sigma\rho}  = \mp \dfrac{1}{\sqrt{|g|}}$,
$\varepsilon^{0123}  = - 1/\sqrt{|g|}$.

According to the definition (\ref{70959108}) of $F_{\mu\nu}$, the set of equations (\ref{35357121}) with (\ref{65713937})
contains four equations for four components $A_{\mu}$.

The energy-momentum conservation law  is written in the Cartesian coordinates of flat space-time in the form
\begin{equation}
\label{71343190}
\frac{\p T^\mu_{.\nu}}{\p x^\mu}  {}={}  0
\;\;,
\end{equation}
where the model energy-momentum tensor is
\begin{equation}
\label{71416389}
T^\mu_{.\nu} \equiv  \left[f^{\mu\rho}\,F_{\nu\rho}  {}-{}
\chi^{-2}\,\left(\L {}-{} 1\right)\,\delta^\mu_\nu\right]/4\pi
\;\;.
\end{equation}

With the definitions
$E_i \equiv F_{i0}$, $B^i \equiv -\varepsilon^{0ijk}\, F_{jk}/2$,
$F_{ij} = \varepsilon_{0ijk}\, B^k$,
$D^i \equiv  f^{0i}$, $H_i \equiv \varepsilon_{0ijk}\, f^{jk}/2$,
$f^{ij} = -\varepsilon^{0ijk}\, H_k$ (Latin indexes take the values $1,2,3$),
 the model equations can be written as
nonlinear equations for the electromagnetic field (see also \cite{Chernitskii2004a}).

This system has the characteristic equation \cite{Chernitskii1998b}
\begin{equation}
\label{CharEq}
\tilde{g}^{\mu\nu}\,\frac{\p \Phi}{\p x^\mu}\,\frac{\p \Phi}{\p x^\nu}=0
\;,
\end{equation}
where $\Phi (x^\mu)=0$ is the equation of a characteristic surface,
and the induced metric $\tilde{g}^{\mu\nu}$ has the following very notable form,
which is specific for the model under consideration:
\begin{equation}
\label{73144072}
\tilde{g}^{\mu\nu} \equiv g^{\mu\nu} - 4\pi\,\chi^2\,T^{\mu\nu}
\;\;.
\end{equation}
As we see, the induced metric $\tilde{g}^{\mu\nu}$ includes the energy-momentum tensor $T^{\mu\nu}$
defined in (\ref{71416389}).

A calculation of the determinant for the effective metric (\ref{73144072})
in Cartesian coordinates of flat space-time gives
\begin{equation}
\label{49991296}
\det{\tilde{g}^{\mu\nu}} = -1
\;\;.
\end{equation}

Using (\ref{71343190}), (\ref{73144072}), (\ref{49991296}) we have the following condition for the
effective metric in Cartesian coordinates $x^\mu$:
\begin{equation}
\label{50668693}
\frac{\p \sqrt{|\tilde{g}|}\,\tilde{g}^{\mu\nu}}{\p x^\mu}  {}={}  0
\;\;,
\end{equation}
where $\tilde{g}=(\det{\tilde{g}^{\mu\nu}})^{-1}=-1$ is the determinant of the inverse tensor
for $\tilde{g}^{\mu\nu}$.

The condition (\ref{50668693}) for the metric plays an important role in gravitational theory.
This condition was considered by V.A.~Fock  in connection with the concept of harmonic coordinates \cite{Fock1959}.
On the significance of this condition see also \cite{LogunovMestvirishvili2000}.

The model equation system can be written in a form explicitly containing the components $A_{\mu}$.
Using the definition (\ref{73144072}), we have, in cartesian coordinates of flat space-time, the following
set of equations \cite{Chernitskii1999}:
\begin{equation}
\label{74643270}
 \left(\tilde{g}^{\mu\sigma}\,\tilde{g}^{\nu\rho} - \tilde{g}^{\mu\rho}\,\tilde{g}^{\nu\sigma}\right)
\frac{\p^2 A_{\rho}}{\p x^\mu \p x^\sigma} {}={} 0
\;\;.
\end{equation}
These equations differ from the corresponding equations of usual linear electrodynamics only by
the substitution $g^{\mu\nu}\to \tilde{g}^{\mu\nu}$.

\section{Electromagnetic interaction}

The electromagnetic interaction appears in this model as an electromagnetic force acting on a massive charged particle \cite{Chernitskii1999}
and a moment of force acting on a particle with an electric or magnetic dipole moment and spin \cite{Chernitskii2006a}.
The appropriate dynamical equations follow from integral conservation laws for the field energy-momentum and
angular momentum, respectively (for details see \cite{Chernitskii1999,Chernitskii2006a}).
These obtained equations, characterizing the electromagnetic interaction, have the corresponding classical form.

In this case, the mass of a particle appears as the full field energy of the appropriate particle solution
in a proper coordinate system.
The spin appears as the full angular momentum of the electromagnetic field for the particle solution
in the proper coordinate system.
There exist static electromagnetic field configurations with spin \cite{Chernitskii1999,Chernitskii2003b}.

The force and the moment of force contain the electromagnetic field of distant particles in the first power.
Thus we can say that the electromagnetic interaction appear in the first order in the small field of
distant particles.

\section{Gravitational interaction}

An explanation of the gravitational interaction in the scope of this model based on the effective Riemannian space
with the metric $\tilde{g}^{\mu\nu}$ (\ref{73144072}) induced by the electromagnetic field.
According to the general method stated in the Introduction, distant particles modify the propagation conditions
for particle under consideration by means of this effective Riemannian space induced
by the field of distant particles. The effective metric includes the electromagnetic field components in
even powers. Thus we can say that the gravitational interaction appear in the second order in the small field of
the distant particles.

The cause of the gravitational interaction in this approach is the energy density of the distant particle
field. But to have the real behaviour of the gravitational potential, i.e. $1/r$, we must take into account
a quick-oscillating part of the distant particle field with an electromagnetic wave background.
In this case, an averaging can give the necessary behaviour of the energy density, $1/r$
(for some details see \cite{Chernitskii2002b,Chernitskii2006b}).

The gravitational constant in this approach is proportional to an amplitude
of the wave background. But this amplitude can have different values for different regions of space.
Consequently, the gravitational constant can be varied in this approach.
In particular, the so-called dark matter effect can have this origin.

\section{Conclusion}

Thus the present approach, based on a consistent application of the idea
of a nonlinear local unified four-vector field, can really unify electromagnetism and gravitation.

\small

%\bibliographystyle{jhep}
%\bibliography{CONFARTICLE0603}

%\end{document}

\end{document}